\documentclass[manuscript]{aastex}

\usepackage{amsmath}
\usepackage{bm}

\usepackage{natbib} 
\usepackage{aas_macros} 
\usepackage{comment}

\newcommand{\bfff}{}
\newcommand{\bff}{}
\renewcommand{\bf}{}

\begin{document}

\title{Temperature Structure of the Pipe Nebula Studied by the Intensity Anomaly of the OH 18 cm Transition}


\author{Yuji Ebisawa}
\affiliation{Department of Physics, The University of Tokyo, Hongo, Bunkyo-ku, Tokyo 113-0033, Japan}
\author{Nami Sakai}
\affiliation{RIKEN, 2-1 Hirosawa, Wako, Saitama 351-0198, Japan}
\author{Karl M. Menten}
\affiliation{Max-Planck-Institut f\"{u}r Radioastronome, Auf dem H\"{u}gel 69, D-53121 Bonn, Germany}
\author{Yoko Oya}
\affiliation{Department of Physics, The University of Tokyo, Hongo, Bunkyo-ku, Tokyo 113-0033, Japan}
\author{Satoshi Yamamoto}
\affiliation{Department of Physics, The University of Tokyo, Hongo, Bunkyo-ku, Tokyo 113-0033, Japan}

\begin{abstract}
We present observations of the four hyperfine structure components of the OH 18 cm transition
(1612, 1665, 1667 and 1720 MHz) toward a filamentary dark cloud, 
the Pipe nebula, 
with the Green Bank Telescope. 
A statistical equilibrium analysis is applied to the spectra,
and the kinetic temperature
of a diffuse molecular gas surrounding dense cores is 
determined accurately; 
the derived temperature ranges from 40 K to 75 K. 
From this result, we assess the heating effect on the filamentary structure of the nebula's ``stem'' region
due to UV photons from a nearby star $\theta$-Ophiuchi and
a possible filament-filament collision in the interface of the ``stem'' and ``bowl'' regions.
In the stem region, the gas kinetic {\bf temperature} is found
to be almost independent of the {\bf apparent} distance from $\theta$-Ophiuchi: 
the UV-heating effect by the star is not {\bf visible.} 
On the other hand, the gas kinetic temperature is raised,
as high as $\sim$75 K, at the interface of the two filamentary structures. 
This result provides us with an additional {\bf support to} 
the filament-filament collision {\bf scenario} in the Pipe nebula. 
\end{abstract}


\section{Introduction}\label{sec:mainabs>intro}
    The 18 cm transition of the hydroxyl radical (OH) has extensively been observed toward
    molecular clouds, late-type stars and external galaxies
    {\bf \citep[e.g.][]{Harju2000, Caswell2004, Hoffman2005, Wolak2012, Darling2002, Ebisawa2015, Ebisawa2019, Li2018, Xu2016a}.} 
    It is the $\Lambda$-type doubling transition in the ground rotational state,
    and consists of the four hyperfine structure (hfs) lines at
    1612, 1665, 1667, and 1720 MHz.
    It is well known that the {\bfff observed} intensity ratio of the four hfs lines {\bfff is} 
    often {\bfff different} from the {\bfff ratio} expected in the local thermodynamic equilibrium (LTE) condition; 
    $I_{\rm 1612}:I_{\rm 1665}:I_{\rm 1667}:I_{\rm 1720}=1:5:9:1$
    \citep[e.g.][]{Harju2000, Ebisawa2015}.
    Such intensity 
    {\bff anomalies can be} 
    caused by
    collisional and/or radiative processes among the {\bff rotational} states
    \citep{Elitzur1976a, Langevelde1995, Ebisawa2019}.
    \citet{Ebisawa2015}, 
    {\bff on the basis of their statistical equilibrium calculations,} 
    reported that the intensity ratio is sensitive
    to the gas kinetic temperature. 
    They demonstrated that the gas kinetic temperature of warm ($T_k$ $>$ 30 K) molecular clouds can be
    determined accurately from the observed intensities.
    Furthermore, the critical density of the OH 18 cm transition is so low ($<$ 10$^2$ cm$^{-3}$) that
    a diffuse {\bfff molecular} gas surrounding {\bfff dense cores,} 
    which is generally difficult to {\bff trace} 
    by rotational transitions of CO
    (i.e. so-called CO-dark gas),
    could be observed by the OH 18 cm transition \citep{Ebisawa2015, Xu2016a, Tang2017}.
    In this paper, this method for temperature determination
    is applied to a nearby molecular cloud, the Pipe nebula.

    The Pipe nebula is a filamentary ($\sim$3 pc $\times$ 18 pc) massive ($\sim$10$^4$ M$_{\odot}$) molecular cloud complex \citep{Frau2015},
    {\bff that} 
    was first identified in the {\bff lines of} CO and its {\bff isotopologues} 
    by \citet{Onishi1999}.
    Figure $\ref{fig:pipe_map}$ shows the 2MASS visual extinction ($A_K$) map
    observed {\bfff for} the whole part of the Pipe nebula \citep{Lombardi2006}.
    As shown in Figure $\ref{fig:pipe_map}$, the Pipe nebula roughly consists of three regions;
    B59 at the western end,
    the stem region {\bfff having} a characteristic filamentary structure,
    and the bowl region exhibiting an extended structure in the eastern part.
    Since the Pipe nebula is a nearby ($\sim$145 pc) object \citep{Alves2007}
    with little star formation activity except for the B59 region \citep{Onishi1999},
    it has been a good target to study molecular-cloud formation \citep{Lada2008, Romanzuniga2010}.
    {\bff Its} 
    magnetic-field structure 
    {\bff has been investigated} 
    by polarimetry \citep{Alves2008, Alves2014, Franco2010}.
    A bright B2 IV star, $\theta$ Ophiuchi (HD 157056), is located at about 3 pc
    away from the Pipe nebula (Figure \ref{fig:pipe_map}),
    {\bff and feedback with this star has been invoked for shaping} 
    the structure and the formation process of the {\bff nebula} 
    \citep{Onishi1999, Gritschneder2012}.

    {\bff The} velocity structure of the Pipe nebula was studied
    in the $^{12}$CO and its isotopologue lines.
    Figure $\ref{fig:pipe_map_CO}$ shows the integrated intensity maps of
    the $^{13}$CO ($J=1-0$) (cyan contours)
    and $^{12}$CO ($J=1-0$) (red contours) lines by \citet{Onishi1999}
    overlaid on the visual extinction ($A_K$) map of Figure $\ref{fig:pipe_map}$.
    The $^{12}$CO map is prepared by integrating over the velocity
    range from 6 to 10 km s$^{-1}$ \citep{Onishi1999}.
    As shown in Figure $\ref{fig:pipe_map_CO}$,
    the $^{13}$CO line traces a filamentary structure elongated along the west to east direction,
    which is seen in the visual extinction map.
    We hereafter call this structure {\bff the} W-E filament.
    On the other hand, the $^{12}$CO map exhibits another filamentary structure
    extending along the south to north direction (S-N filament)
    which is perpendicular to the W-E filament.
    According to \citet{Frau2015},
    the W-E filament has a blue-shifted velocity ($V_{\rm LSR}$ of 2--4 km s$^{-1}$),
    whereas the S-N filament has a red-shifted velocity ($V_{\rm LSR}$ of 6--7 km s$^{-1}$).
    These two filamentary structures {\bff overlap} 
    on the bowl region.
    On the basis of their combined analyses of the optical polarimetry,
    the visual extinction, and $^{13}$CO data, 
    {\bfff \citet{Frau2015} suggested that these two structures} 
    are colliding with each other in the overlapping region.
    They found that {\bff the} gas in the bowl region tends to show a higher density and a broader linewidth,
    as well as a stronger polarization degree
    and a smaller dispersion of polarization angle
    in the optical observations. 
    {\bff These results are discussed in terms of} 
    compressive motion caused by {\bff a} filament-filament collision.

    In this study,
    we investigate the {\bfff nature} of the Pipe nebula by {\bfff focusing on} the following two effects:
    (1) 
    UV heating from $\theta$-Oph,
    and (2) 
    collisions between the W-E and S-N filaments.
    For this purpose, we explore a temperature structure of the Pipe nebula
    by observing the OH 18 cm transition.
    If the Pipe nebula {\bff were} 
    affected by 
    UV heating from $\theta$-Oph,
    the gas kinetic temperature {\bfff would} 
    {\bff increase} with a decreasing distance from $\theta$-Oph.
    In addition, the temperature {\bfff would} be raised at the interface of the two filaments,
    if {\bff a} filament-filament collision {\bff occurred.} 
    We examine these two effects on the basis of our statistical equilibrium calculations
    for the OH 18 cm transition.

\section{Observation}\label{sec:mainabs>observation}
    We observed the four hfs components of the OH 18 cm transition toward the Pipe nebula
    with the Robert C. Byrd Green Bank Telescope (GBT) in 2018 and 2019.
    {\bff 
    Four hfs lines of the OH 18 cm transitions were observed with the L-Band receiver. 
    The beam size (FWHM) is 470\arcsec, 
    where the beam efficiency is 82\%. 
    A typical system temperature was from 18 to 29 K. 
    The resolution of the backend correlator was tuned to be 0.26 kHz, 
    corresponding to the velocity resolution of 0.047 km s$^{-1}$ at 1.65 GHz.} 
    The integration time was about 1.5 hours/position.
    Such a long integration time was necessary to obtain {\bff an} rms noise temperature
    of $\sim$10 mK at a velocity resolution of 0.2 km s$^{-1}$.
    
    The {\bff observations were} conducted 
    {\bff toward the positions indicated by the yellow circles arranged in the three strips shown in Figure \ref{fig:pipe_map_CO}.} 
    The reference (0$'$, 0$'$) positions of strip-1 and strip-2 are
    ($\alpha_{2000}$, $\delta_{2000}$) = (17$^{\rm h}$20$^{\rm m}$49$^{\rm s}$.0, -26$^\circ$53$'$8.0$''$)
    and
    (17$^h$27$^m$12$^s$.0, -26$^\circ$42$'$59.0$''$), respectively.
    {\bff The observations along strip-1 are arranged to assess the impact of UV heating} 
    from $\theta$-Oph
    by determining the gas kinetic temperature
    as a function of the distance from the star.
    The strip-2 is parallel to the W-E filament
    seen in the visual extinction ($A_K$) map (Figure $\ref{fig:pipe_map}$)
    and the $^{13}$CO map (cyan contours in Figure $\ref{fig:pipe_map_CO}$).
    With the observation along strip-2,
    we examine 
    {\bff whether the material at the interface} 
    of the W-E and S-N filaments 
    {\bff has an increased temperature.} 
    {\bff Strip-3} is perpendicular to strip-2 
    {\bff and is centered on a 
    position 
    located} on the interface of the filaments, 
    {\bff i.e. at the offset of (32$'$, 0$'$) in strip-2.} 

    Intensities of the radio continuum background emission are evaluated for each observed position
    from the HIPASS 1.4 GHz continuum data \citep{Calabretta2014} by following the method {\bff of} 
    \citet{Tang2017}.
    The intensities 
    {\bff along} strip-1 and strip-2 are {\bff found} 
    to be
    $\sim$4.5--4.9 K and $\sim$5.1--5.8 K at 1667 MHz, respectively,
    and tend to be {\bfff higher} toward the Galactic center {\bff direction.}


\section{Result}\label{sec:mainabs>result}
    Figure $\ref{fig:pipe_spectra_strip1}$ shows the spectra of the OH 18 cm transition
    observed along strip-1 (Figure $\ref{fig:pipe_map_CO}$).
    The 1612 MHz line shows a clear absorption feature for all the observed positions.
    In contrast, the 1720 MHz line shows a brighter emission
    than expected {\bfff for} the LTE condition (see Section $\ref{sec:mainabs>intro}$).
    This trend is the same as that seen in warm diffuse clouds \citep{Ebisawa2015}.
    Hence, we are looking at a warm ($>$ 40 K) gas component of the Pipe nebula
    in the OH 18 cm transition.
    The 1665 and 1667 MHz lines are the strongest at the central position,
    and they become weaker to cloud peripheries.
    {\bf The peak intensity, the line width, and the LSR velocity 
    are derived by fitting the Gaussian function to each observed spectrum, 
    where the line width and the LSR velocities are assumed to be identical 
    among the hyperfine components. 
    The results are summarized in Table \ref{table:pipe_linepara_strip1}.}     
    The {\bff average} 
    LSR velocity of 3.3 km s$^{-1}$ is shown
    by blue dotted lines in Figure $\ref{fig:pipe_spectra_strip1}$. 
    {\bff This LSR velocity is consistent with that reported by \citet{Onishi1999}.} 

    Figures $\ref{fig:pipe_spectra_strip2_stem}$ and $\ref{fig:pipe_spectra_strip2_bowl}$
    show the spectra of the OH 18 cm transition observed along strip-2 (Figure $\ref{fig:pipe_map_CO}$)
    in the stem ($\Delta\alpha=-16', -8', 0', 8', 16', 24'$)
    and bowl ($\Delta\alpha=32', 40', 48', 56', 64', 72', 80'$) regions, respectively.
    Toward the stem region (Figure $\ref{fig:pipe_spectra_strip2_stem}$),
    the absorption {\bff in} the 1612 MHz line and the enhanced emission of the 1720 MHz line are observed,
    as is the case {\bff for} strip-1 (Figure $\ref{fig:pipe_spectra_strip1}$).
    The main lines (1665 and 1667 MHz) are observed in emission
    at $V_{\rm LSR}$ {\bf of ($3.5-4.0$) 
    km s$^{-1}$} in this region, 
    {\bf except for the ($+24'$, $0'$) position.} 
    However, the main lines sharply turn from emission to absorption
    around the boundary between the stem region and the bowl region.
    This sharp change is apparent by comparing the spectra at the (+16$'$, 0$'$) and (+24$'$, 0$'$) positions.
    Namely, the main lines show emission at the (+16$'$, 0$'$) position,
    whereas clear absorption features of these lines are seen in the (+24$'$, 0$'$) position.
    {\bff Absorption} in the main lines {\bff is} present in {\bf almost} all the observed positions
    in the bowl region (Figure $\ref{fig:pipe_spectra_strip2_bowl}$).
    It should be noted that the radio continuum background emission in the bowl region is brighter
    than that in the stem region \citep{Calabretta2014},
    which might explain the absorption of the main lines in the bowl region.
    However, the contribution of the background emission
    cannot fully explain the sharp transition from emission to absorption of the main lines
    at the interface of filaments around (24$'$, 0$'$) or (32$'$, 0$'$) positions.
    {\bff Later,} in Section $\ref{sec:mainabs>strip2}$,
    we {\bff argue that a} 
    rise in the gas kinetic temperature produces the absorption at the interface.

    In addition, several positions in the bowl region ($\Delta\alpha$ = 56$'$, 64$'$ (marginal), 72$'$)
    have {\bff a} velocity component ($V_{\rm LSR}$ $\sim${\bf ($5-6$) km s$^{-1}$)} {\bff that shows} 
    the 
    absorption {\bff in} the 1720 MHz line.
    According to \citet{Ebisawa2019},
    {\bf the} 1720 MHz absorption traces relatively cold ($T_k$ $<$ 30 K) {\bf gas with high OH column density ($>$ 10$^{15}$ cm$^{-2}$)} 
    illuminated by FIR radiation from dust grains 
    {\bfff in a surrounding warm cloud.}
    The positions showing the above {\bff characteristics} are indeed located
    in the inner part of the S-N filament (Figure $\ref{fig:pipe_map_CO}$).
    This suggests that the 
    component {\bf showing the 1720 MHz absorption} 
    {\bff represents} 
    {\bff a} relatively cold part of the filament.
    In addition, the 1720 MHz absorption feature is seen only in the eastern positions of the bowl region.
    According to the {\bff archival} IRAS 
    and Herschel 
    continuum maps, the dust temperature is higher toward the eastern part of strip-2.
    {\bfff Since} FIR radiation 
    at {\bfff the} wavelengths of the 53 and 35 $\mu$m 
    {\bfff plays an important role in {\bff causing} the 1720 MHz absorption \citep{Ebisawa2019},} 
    {\bfff the} FIR pumping {\bfff effect becomes} more efficient with a higher 
    {\bf dust} temperature 
    {\bfff of the surrounding {\bf cloud.}} 
    {\bff This qualitatively explains the} 
    appearance of the 1720 MHz absorption 
    {\bff in the eastern part.} 

    As shown in Figure $\ref{fig:pipe_spectra_strip2_stem}$, the four hfs components of the OH 18 cm transition
    {\bff show} 
    two velocity components in the stem region at $V_{\rm LSR}$ of $\sim$3.0 and $\sim$4.0 km s$^{-1}$
    (blue and red dotted lines, respectively).
    In contrast, the 3 km s$^{-1}$ component is generally faint or almost absent in the bowl region
    (Figure $\ref{fig:pipe_spectra_strip2_bowl}$).
    whereas the $\sim$4.0 and $\sim$5.0--6.0 km s$^{-1}$ components are 
    observed there. 
    {\bf 
    The line parameters for the two velocity components in the stem and bowl regions are 
    determined by fitting the double Gaussian function to the observed spectra of the four hyperfine components except for the (24, 0), (32, 0), (40, 0), and (72, 0) positions, 
    where the line width and the LSR velocity are assumed to be identical among the hyperfine components. 
    For the (24,0) and (32,0) positions, 
    which {\bf correspond to} the interface of the W-E and S-N filaments, 
    we cannot determine the line parameters assuming two velocity components due to heavy blending.  
    Hence, the line parameters are determined by the single Gaussian fit. For the (72, 0) position, 
    we employ the triple Gaussian fitting. 
    On the other hand, 
    the spectrum of the (40, 0) position is so complicated 
    that we cannot fit the hyperfine structure in any way. 
    The obtained line parameters are summarized in Table \ref{table:pipe_linepara_strip2}.}

    Figure $\ref{fig:pipe_spectra_strip3}$ shows the spectra of the OH 18 cm transition
    observed along strip-3 (Figure $\ref{fig:pipe_map_CO}$).
    The main lines {\bfff appear in} absorption at all the observed positions.
    The two velocity components at $V_{\rm LSR}$ $\sim$3 and 4 km s$^{-1}$ are clearly seen {\bff toward} the southern positions
    (32$'$, -8$'$) and (32$'$, -16$'$),
    whereas they are blended in the (32$'$, 0$'$) position.
    As for the (32$'$, 8$'$) position,
    the two components (4.16 and 5.84 km s$^{-1}$) are seen.
    The double Gaussian fit is applied to {\bfff strip-3} positions except for the (32$'$, 0$'$) position.
    The obtained line parameters are summarized in Table $\ref{table:pipe_linepara_strip3}$.

\section{Heating Effect from $\theta$-Ophiuchi}\label{sec:mainabs>strip1}
    {\bff 
    We perform a least-square fit on the hfs 
    {\bf intensities derived above to} 
    determine the gas kinetic temperature and the OH column density for all
    the observed positions along strip-1 (Figure $\ref{fig:pipe_map_CO}$) 
    {\bf by using} 
    the statistical equilibrium calculation described by \citet{Ebisawa2015}. 
    {\bf Here, we assume} 
    {\bff an} H$_2$ column density of 10$^3$ cm$^{-3}$ and {\bff an} H$_2$ ortho-to-para ratio of 3
    (Table $\ref{table:pipe_Tk_strip1}$).
    Since the hfs {\bf intensities} 
    of the OH 18 cm transition {\bf are} 
    insensitive to {\bff an} H$_2$ density from 10$^2$ cm$^{-3}$ to 10$^6$ cm$^{-3}$ \citep{Ebisawa2015},
    the assumption of {\bff an} H$_2$ density of 10$^3$ cm$^{-3}$
    is just arbitrary, and does not affect the results.
    Here, the intensity of the background continuum emission is evaluated from the HIPASS 1.4 GHz continuum data
    \citep{Calabretta2014}, as {\bfff mentioned} in Section $\ref{sec:mainabs>observation}$.
    The effects of the FIR radiation and the line overlap are not included in this calculation.
    \added{The collisional rate coefficients of OH reported by \citet{Offer1994} are employed in the calculations. }
    {\bf In this analysis, 
    the hfs intensities including the 1612 MHz absorption 
    are well reproduced, 
    and the gas kinetic temperature 
    and the OH column density at each position are determined.} 

    Figure $\ref{fig:pipe_Tk_strip1}$ shows the derived gas kinetic temperatures
    as a function of the angular offset along strip-1 (Figure $\ref{fig:pipe_map}$), 
    {\bf where the} 
    {\bff star} $\theta$-Ophiuchi is closer to the {\bff northeastern} positions 
    {\bf on the plane of the sky.} 
    {\bf 
    The gas temperature is found to be slightly higher at the cloud peripheries than at the cloud center along the strip 1. 
    Since the strip 1 is well apart from the interaction region between the stem and bowl structure, 
    the main heating mechanism is {\bf likely} the UV heating. 
    Here, we qualitatively discuss its effect on strip 1. 
    According to Onishi et al. (1999), the peak H$_2$ column density is $8.5 \times 10^{21}$ cm$^{-2}$, 
    which corresponds to the visual extinction of about 8.5. 
    However, the cloud peripheries are illuminated by external UV radiation, 
    and enhancement of the gas kinetic temperature there is expected by photodissociation region (PDR) models 
    \citep[e.g.,][]{TielensHollenbach1985, Spaans1996}. 
    Such a PDR effect is indeed reported for the $\rho$-Oph cloud in the OH 18 cm transition 
    \citep{Ebisawa2015}. 
    Hence, the above temperature structure can be understood in terms of the external UV radiation effect. 
    On the other hand, we find no temperature gradient as a function of the apparent distance from $\theta$-Oph. 
    This result suggests that the heating effect from $\theta$-Oph 
    {\bf is not seen in the observed temperature structure.}  
    {\bf Hence, the heating source would likely be the interstellar UV radiation. 
    It should be noted, however, that} 
    we also expect no temperature gradient along the E-W direction due to the illumination by $\theta$-Oph, 
    {\bf if} $\theta$-Oph is illuminating the observed filament from the back side or the front side. 
    } 

\section{The Mechanism Causing Absorption in the Main Lines}\label{sec:mainabs>strip2}
    In this section, we explore the origin of the absorption {\bff observed in} 
    the main {\bff OH hfs} lines {\bfff (1665 and 1667 MHz)} of the OH 18 cm transition,
    which is observed in the bowl region of the Pipe nebula
    (Figures $\ref{fig:pipe_spectra_strip2_stem}$ and $\ref{fig:pipe_spectra_strip2_bowl}$).
    {\bfff The} absorption is exclusively observed in the bowl region
    where these filaments {\bff overlap,} 
    and {\bff it} starts to appear from the interface of the filaments
    (i.e. (24$'$, 0$'$) position) to the bowl region.
    {\bfff Hence, 
    the main line absorption 
    might be related to the interaction between the W-E and S-N filaments.}

    Figure $\ref{fig:LVG_mainabs_Tk}$ shows the intensities of the four hfs lines of the OH 18 cm transition
    as a function of the gas kinetic temperature, 
    {\bf which are derived by using the statistical equilibrium calculation \citep{Ebisawa2015}.} 
    The H$_2$ density, the H$_2$ ortho-to-para ratio, the OH column density
    and the intensity of the radio continuum background emission are
    assumed to be 10$^3$ cm$^{-3}$, 3, 10$^{14}$ cm$^{-2}$ and 5 K, respectively. 
    {\bf No effect of the FIR radiation is included.} 
    The 1665 and 1667 MHz lines appear in absorption
    with a gas kinetic temperature higher than about 60 K.
    The absorption becomes deeper for a higher gas kinetic temperature.
    This prediction suggests that the absorption of the main lines observed in the bowl region
    traces warm gas with {\bff a} gas kinetic temperature higher than 60 K.
    It should be noted that 
    {\bff the fainter the continuum background temperature, 
    the higher a gas kinetic temperature is required to reproduce the absorption.} 
    If the cosmic microwave background (2.73 K) 
    {\bff were to provide the continuum background,} 
    {\bff a} gas kinetic temperature higher
    than about 100 K would be necessary to reproduce the absorption.

    
    {\bff Absorption in the 18 cm main hfs lines of OH} 
    can qualitatively be explained in terms of the {\bfff following} non-LTE effect.
    Most of the OH molecules excited to the rotationally excited states
    in the $^2\Pi_{3/2}$ ladder ($J$ = 5/2, 7/2 or higher) by collisions with H$_2$ molecules
    are quickly de-excited to the ground rotational state ($^2\Pi_{3/2}$ $J$ = 3/2)
    {\bfff through} rotational transitions within the $^2\Pi_{3/2}$ ladder
    (red downward arrows in Figure $\ref{fig:OH_energy_mainabs}$).
    During the radiative decay, the $\Lambda$-type transitions within each rotational level
    can occur, as represented by the blue and green arrows in Figure $\ref{fig:OH_energy_mainabs}$.
    Thus the populations of the lower $\Lambda$-type doubling levels relative to {\bff those of the upper levels} 
    are slightly increased in the $^2\Pi_{3/2}$ $J$ = 5/2, 7/2, and higher states,
    as represented by the red ellipses in Figure $\ref{fig:OH_energy_mainabs}$.
    These overpopulations in the lower $\Lambda$-type doubling levels subsequently
    produce a similar anomaly in the ground rotational state
    via rotational transitions within the $^2\Pi_{3/2}$ ladder
    (red downward arrows in Figure $\ref{fig:OH_energy_mainabs}$).
    In this situation, the excitation temperatures of the 1665 and 1667 MHz lines are decreased,
    which {\bff causes} absorption against the background radiation.
    
    It should be noted that the $\Lambda$-type transitions in the rotationally excited states
    rarely occur compared to the rotational transitions,
    since Einstein's A coefficients of these transitions are about 10$^9$--10$^{10}$ times
    lower than those of the rotational transitions \citep{Offer1994}.
    This suggests that the $\Lambda$-type transitions in the rotationally excited states
    affect the {\bfff populations} only slightly.
    Nevertheless, such a slight effect on the population can significantly
    change the excitation temperature of the OH 18 cm transition
    because of the small energy gap 
    between {\bfff the} $\Lambda$-type doubling levels ($\Delta E$ $\sim$0.08 K).
    {\bff This} energy gap 
    is so small that the populations of the upper and lower states
    of this transition are almost comparable even in the low temperature.
    For example, assuming {\bff an} excitation temperature of 2 K, 10 K and 100 K,
    the population difference between the upper and lower states
    are about 2\%, 0.4\% and 0.04\%, respectively.
    This implies that the excitation temperature can be lower than the continuum {\bff temperature,} 
    just by raising the lower state population by only one percent or less,
    which would be possible through the $\Lambda$-type transitions. 
    {\bff Collisional excitation} to the upper rotationally excited states
    {\bff becomes} more efficient with increasing gas kinetic temperature,
    which leads to deeper absorption features of the main lines,
    as shown in Figure $\ref{fig:LVG_mainabs_Tk}$.
    Note that the contributions from the $^2\Pi_{1/2}$ state
    {\bfff are not important, because} 
    inter-ladder radiative transitions from $^2\Pi_{1/2}$ to $^2\Pi_{3/2}$ are
    less frequent than those within the $^2\Pi_{1/2}$ or $^2\Pi_{3/2}$ ladder.

    Figure $\ref{fig:LVG_mainabs_nH2}$ shows the expected intensities of the hfs lines of the
    OH 18 cm transition derived from our statistical equilibrium calculations
    as a function of the H$_2$ density.
    Here, the OH column density, the gas kinetic temperature, and the H$_2$ ortho-to-para ratio
    are assumed to be 10$^{14}$ cm$^{-2}$, 75 K and 3, respectively.
    The intensity of the radio continuum background emission is assumed to be 5 K
    as a representative value.
    The absorption features of the main lines as well as {\bff of} the 1612 MHz line
    are reproduced in the H$_2$ density range from 1 to 10$^6$ cm$^{-3}$.
    Moreover, intensities of all the four hfs lines are insensitive to the H$_2$
    density from 10$^2$ to 10$^{6}$ cm$^{-3}$,
    as {\bff was found in} our previous paper \citep{Ebisawa2015}.
    This suggests that the radiative transitions are dominant over the collisional transitions
    at this H$_2$ density range {\bff for a} gas kinetic temperature of 75 K.
    This is consistent with the above picture on the origin of the main lines absorption,
    where the absorption is essentially produced by the $\Lambda$-type transitions
    in the rotationally excited states during the rotational cascade.

    In order to examine if our statistical equilibrium analysis
    can quantitatively reproduce the absorption feature in the main lines
    as well as the intensities of the 1612 and 1720 MHz satellite lines,
    we first perform {\bff a} least-squares fit on the {\bf intensities of the} OH spectra 
    observed on (+32$'$, 0$'$) position (Figure $\ref{fig:pipe_spectra_fit}$),
    which shows the deepest absorption feature.
    In this analysis, we conduct the same statistical equilibrium calculation
    as that used in Section $\ref{sec:mainabs>strip1}$,
    assuming {\bff an} H$_2$ density of 10$^3$ cm$^{-3}$ and {\bff an} H$_2$ ortho-to-para ratio of 3.
    As a result, the observed intensities of the four hfs lines are well reproduced by the simultaneous fit,
    as shown in the red lines in Figure $\ref{fig:pipe_spectra_fit}$.
    The gas kinetic temperature and the OH column density are determined to be
    73.6 $\pm$ 0.8 K and (5.3 $\pm$ 0.2) $\times$ 10$^{13}$ cm$^{-2}$, respectively.
    The gas kinetic temperature of {\bfff 73.6} K is apparently higher than that
    derived in the stem region along strip-1 (40--60 K) (Figure $\ref{fig:pipe_Tk_strip1}$).

\section{Enhanced Temperature in the Interacting Region}\label{sec:mainabs>temperature}
    We then perform the same least-squares fit on 
    {\bf the hfs intensities of} all the spectra along strip-2
    (Figures $\ref{fig:pipe_spectra_strip2_stem}$ and $\ref{fig:pipe_spectra_strip2_bowl}$)
    with our statistical equilibrium calculations.
    The density and the ortho-to-para ratio of H$_2$ are assumed to be
    10$^3$ cm$^{-3}$ and 3, respectively {\bf in} the fit.
    {\bf An effect of the FIR radiation is ignored, 
    and hence,} 
    the fitting is not conducted for the velocity components showing the 1720 MHz absorption or its faint emission,
    that is the $\sim$5.0 km s$^{-1}$ component of the (48$'$, 0$'$), (56$'$, 0$'$) and (64$'$, 0$'$) positions,
    as well as the 5.7 km s$^{-1}$ component of the (72$'$, 0$'$) position.
    As presented by \citet{Ebisawa2019},
    accurate information on the FIR field and the cloud structure are necessary
    to analyze the 1720 MHz absorption.
    Due to the lack of the information {\bff for the Pipe nebula,} 
    we exclude these components from the present analysis.
    In addition, the gas kinetic temperature cannot be determined for the (80$'$, 0$'$) position,
    {\bf because of poor signal-to-noise ratios of the satellite lines.} 

    The derived gas kinetic temperatures and OH column densities are summarized in Table $\ref{table:pipe_Tk_strip2}$.
    Figure $\ref{fig:pipe_Tk_strip2}$ shows {\bff the} 
    gas kinetic temperatures
    as a function of the angular offset in right ascension from the (0$'$, 0$'$) position.
    As for the stem region (right five points in Figure $\ref{fig:pipe_Tk_strip2}$),
    the temperatures of the $\sim$3 and $\sim$4 km s$^{-1}$ components are determined to be
    about 40--60 K, as represented by blue and red colors, respectively.
    The temperature is slightly increased at the (24$'$, 0$'$) position,
    and reaches the highest value of 73.6 $\pm$ 0.8 K at the interface of the stem and bowl regions.
    (i.e. the (32$'$, 0$'$) position).
    Note that the quoted errors do not include the systematic uncertainty
    due to the collisional rates, radiative transfer {\bff effects,} 
    and the {\bff fact that we are} 
    neglecting the FIR radiation and line overlap.
    Toward the eastern positions in the bowl region (left four points in Figure $\ref{fig:pipe_Tk_strip2}$),
    the gas kinetic temperatures are determined to be 50--60 K,
    which are comparable to those determined for the stem region ($\Delta\alpha$ = -16$'$--16$'$).
    As shown in Figure $\ref{fig:pipe_Tk_strip2}$, the derived gas kinetic temperature
    is indeed {\bfff highest} at the positions located on the interface of the two filamentary structures (Figure $\ref{fig:pipe_map_CO}$)
    ($\Delta\alpha=24', 32'$) (Figure $\ref{fig:pipe_map_CO}$).
    The rise of the gas kinetic temperature can be interpreted
    as the {\bff effect of} heating {\bff caused} by the collisions between the filaments.

    Similar least-square analyses {\bf for the hfs intensities} are performed for the spectra observed along strip-3,
    as shown in Figure $\ref{fig:pipe_Tk_strip3}$.
    The results are summarized in Table $\ref{table:pipe_Tk_strip3}$.
    {\bff Gas} kinetic temperatures of the $\sim$4 km s$^{-1}$ component are
    about 70 K for the four positions,
    whereas they are $\sim$60 K for the $\sim$3 and 5 km s$^{-1}$ components.
    Since all these positions are located {\bff at} the interface of the filaments
    (Figure $\ref{fig:pipe_map_CO}$),
    the high gas kinetic temperature ($\sim$70 K) compared to the
    western and eastern positions on strip-2 ($\sim$50--60 K)
    further supports the existence of the filament-filament collision.
    
    {\bf 
    When the collision velocity is assumed to be the velocity difference between the stem and bowl region ($\sim$1.5 km s$^{-1}$), 
    the kinetic energy released by the collision is estimated to be $\sim 10^{-10}$ erg cm$^{-3}$, 
    where the H$_2$ density of $10^{3}$ cm$^{-3}$ is arbitrarily assumed. 
    This can be balanced by the cooling rate by 
    \citet{GoldsmithLanger1978}, 
    and the equilibrium temperature is roughly estimated to be 100 K. 
    Here, we assume the cooling time of $10^4$ year according to 
    \citet{Inoue2012}. 
    This temperature is comparable with that observed in the interacting region. 
    Thus, this rough estimate implies 
    that the slight temperature enhancement is possible in the interacting region. 
     }

    {\bff Warm} gas {\bff with a temperature of} 74 K, 
    heated by the {\bfff filament-filament collision,} 
    is cooled down mainly by {\bff rotational line emission} 
    of CO 
    for {\bff an} H$_2$ density of 10$^3$ cm$^{-3}$ \citep{Inoue2012}.
    {\bf As mentioned above, 
    the} cooling time ($t_{\rm cool}$) is estimated to be around 10$^4$ year
    according to \citet{Koyama2000} with a cooling rate
    of 10$^{-26}$ erg cm$^{-3}$ s$^{-1}$ \citep{Inoue2012}.
    Assuming that {\bff this} warm gas expands with a constant velocity of 1.5 km s$^{-1}$,
    which corresponds to {\bff the} typical velocity difference between
    the blue-shifted and red-shifted components of the Pipe nebula
    of 3.0 km s$^{-1}$ and 4--5 km s$^{-1}$, respectively,
    the gas can extend in space by about 0.015 pc within the cooling time.
    This length corresponds to a typical spatial scale of a shocked warm region,
    and hence we hereafter denote it as $L_{\rm warm}$ (= 0.015 pc).
    On the other hand, 
    {\bff the extent of the cloud along the line-of-sight} ($L_{\rm LOS}$) 
    can roughly be estimated from the H$_2$ density ($n$(H$_2$)), the OH column density ($N$(OH))
    and the OH fractional abundance ($X$(OH)) {\bf relative to H$_2$} 
    with the following equation:
    \begin{eqnarray}
        L_{\rm LOS} = \frac{N({\rm OH})}{X({\rm OH}) n(\rm{H_2})}.
    \end{eqnarray}
    We employ {\bff an} H$_2$ density of 10$^3$ cm$^{-3}$ and
    {\bff an} OH column density of 5.3 $\times$ 10$^{13}$ cm$^{-2}$,
    {\bff a latter of which} {\bf is} determined for the (0$'$, 32$'$) position (Table $\ref{table:pipe_Tk_strip2}$).
    Assuming {\bff an} OH abundance of 10$^{-7}$,
    which is a typical value in diffuse clouds \citep{Wiesemeyer2012},
    $L_{\rm LOS}$ is estimated to be 
    {\bf $\sim$0.2 pc.} 
    If the OH fractional abundance is one order of magnitude higher ($X$(OH) = 10$^{-6}$),
    which could be possible in a shocked gas according to \citet{Draine1986},
    $L_{\rm LOS}$ could be 0.01 pc.
    The latter value ($L_{\rm LOS}$ {\bf $\sim$0.02} pc) for the shocked gas case ($X$(OH)=10$^{-6}$)
    is comparable to the $L_{\rm warm}$ of 0.015 pc.
    This result suggests the presence of 
    {\bff shock-induced compression} 
    at the interface of the two filaments,
    and supports the filament-filament collisions picture in the Pipe nebula.

\section{Summary}
    We {\bfff have} observed the four hfs components of the OH 18 cm transitions toward the Pipe nebula
    along the two strip lines 
    and {\bfff determined} the gas kinetic temperatures accurately.
    Along strip-1,
    the gas kinetic temperature is found to have no relation
    to the {\bf apparent} distance to the nearby star, $\theta$-Oph.
    This result suggests that the heating effect from $\theta$-Oph 
    {\bff does not significantly contribute to the heating of the Pipe nebula,} 
    {\bf although a possibility that $\theta$-Oph illuminates 
    the cloud from the front side or the back side in the 3D configuration remains.} 
    On strip-2 and {\bfff strip-3,} the absorption feature of the 1665 and 1667 MHz
    main lines is observed in the bowl region,
    and the deepest absorption is observed
    just at the interface of the W-E and S-N filamentary structures.
    Our statistical equilibrium calculations successfully reproduce this absorption feature
    with a gas kinetic temperature higher than $\sim$60 K,
    {\bff where} 
    {\bff an increasing gas kinetic temperature would deepen the absorption.} 
    The derived temperature is indeed the highest at the interface of the filaments,
    indicating 
    {\bff shock heating at the filament interface.} 
    This observation demonstrates a unique characteristic of the OH 18 cm transition
    as a new tool to study the molecular-cloud formation.

\acknowledgements
{\bf The authors thank the anonymous reviewer for valuable comments.} 
The authors are grateful to the staff of the Green Bank Telescope for excellent support. 
This study is financially supported by Grant-in-Aid from Ministry of Education Sports, Science, and Technologies of Japan 
{\bff (25400223, 18H05222, 19H05069, and 19K14753).}

\clearpage
\bibliography{cites.bib}

\clearpage
\begin{figure}[!htbp]
    \centering
    \includegraphics[width=0.7\hsize]{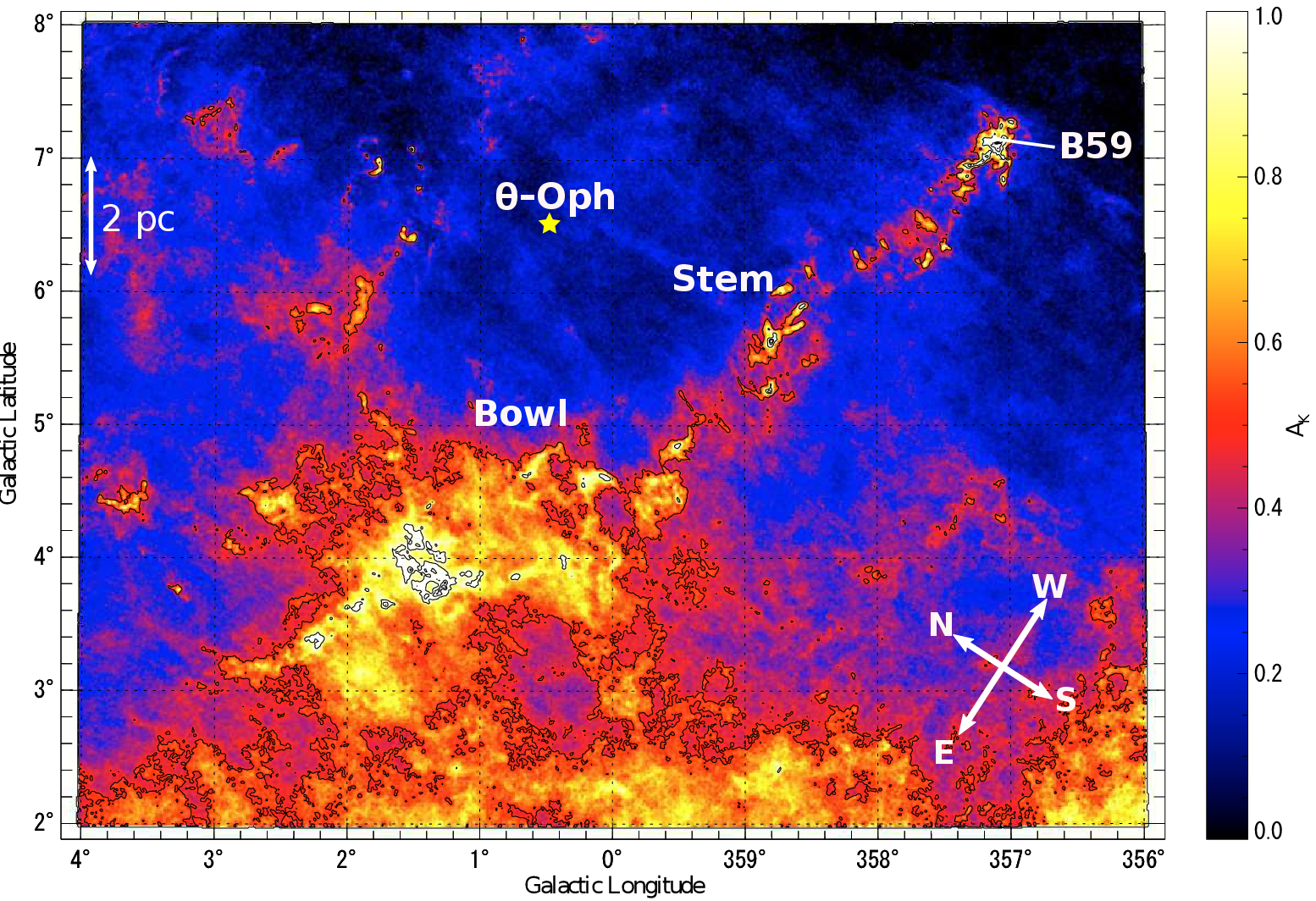}
    \caption{%
        The visual extinction ($A_k$) map observed for the Pipe nebula \citep{Lombardi2006}.
        The position of the $\theta$-Ophiuchi is represented by a yellow star mark.
        \label{fig:pipe_map}}
\end{figure}
\clearpage
\begin{figure}[!htbp]
    \centering
    \includegraphics[width=\hsize]{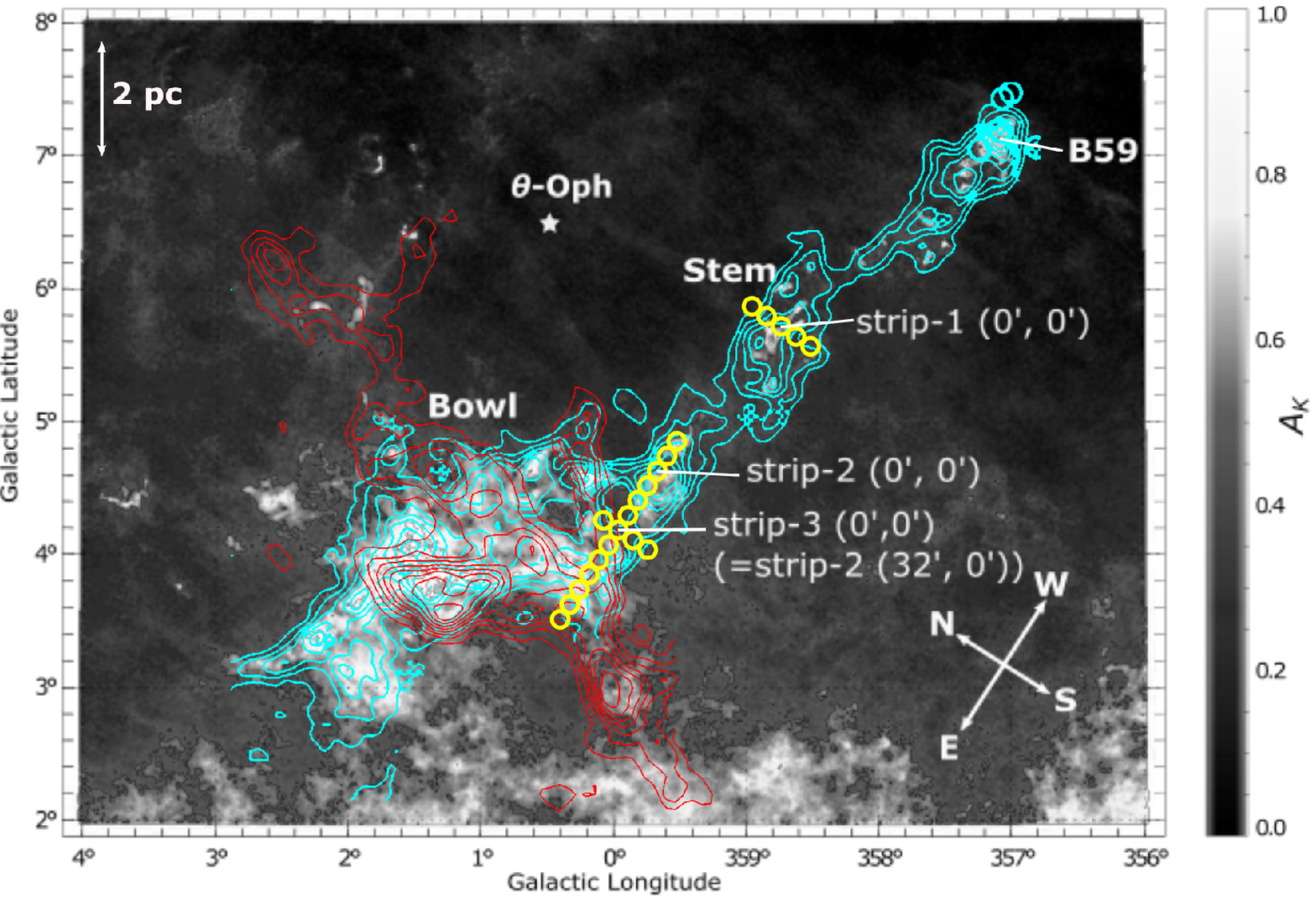}
    \caption{%
        (Cyan)
        The integrated intensity map of the $^{13}$CO ($J$=1--0) emission
        observed by \citet{Onishi1999}.
        (Red)
        The integrated intensity map of the $^{12}$CO ($J$=1--0) emission
        from 6 to 10 km s$^{-1}$ observed by \citet{Onishi1999}.
        (Gray)
        The visual extinction ($A_k$) map reported by \citet{Lombardi2006},
        which is the same as Figure $\ref{fig:pipe_map}$.
        Yellow circles represent the observed positions in the OH 18 cm transition.
        A diameter of the circle corresponds to the HPBW beam size of the GBT
        of 8 arcmin.
        \label{fig:pipe_map_CO}}
\end{figure}
\clearpage
\begin{figure}[!htbp]
    \centering
    \includegraphics[width=\hsize]{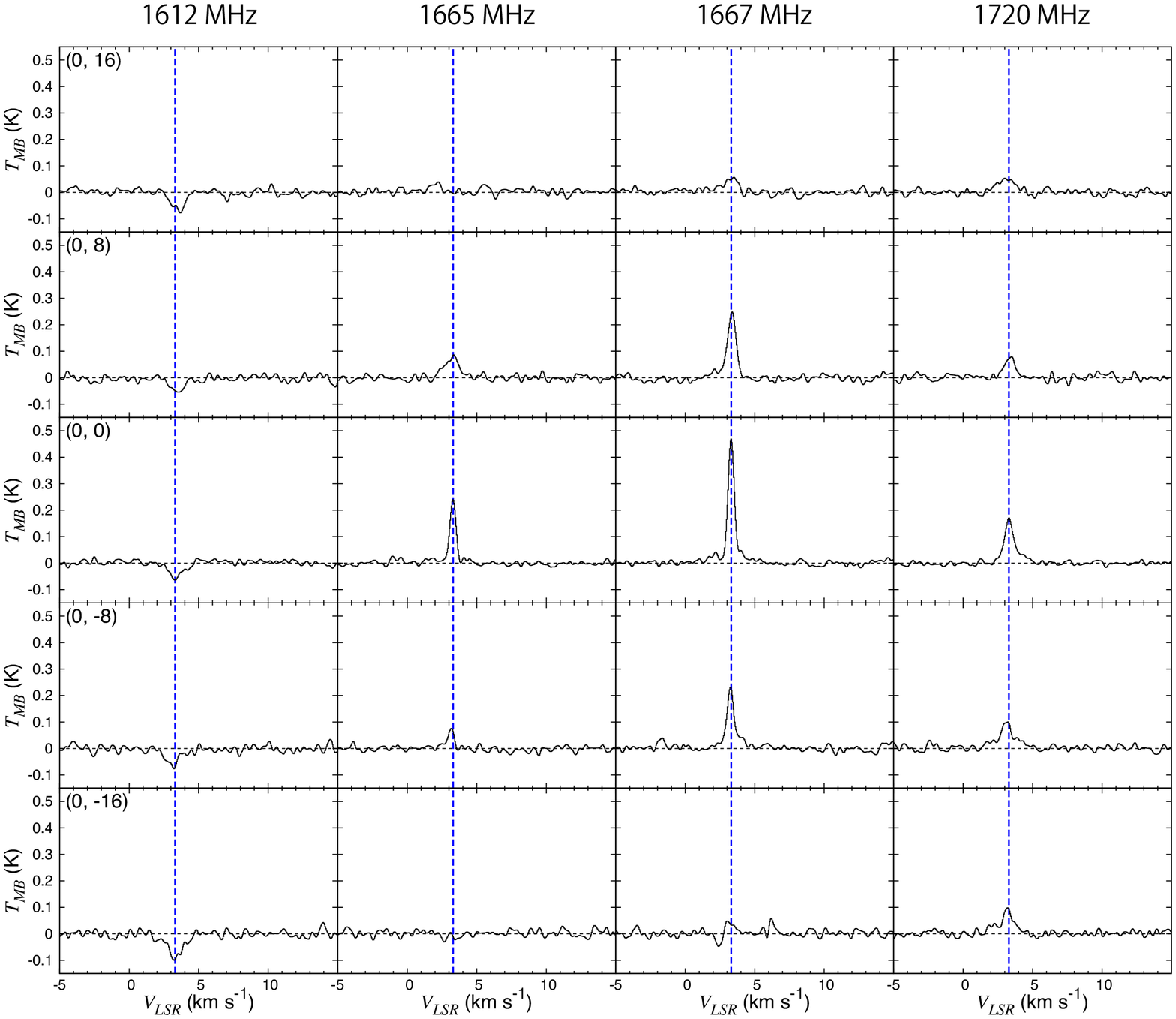}
    \caption{%
        Spectra of the four hfs components of the OH 18 cm transition
        observed toward the Pipe nebula along strip-1 (Figure $\ref{fig:pipe_map_CO}$).
        The blue dotted lines represent the $V_{\rm LSR}$ of 3.3 km s$^{-1}$.
        \label{fig:pipe_spectra_strip1}}
\end{figure}
\clearpage
\begin{figure}[!htbp]
    \centering
    \includegraphics[width=\hsize]{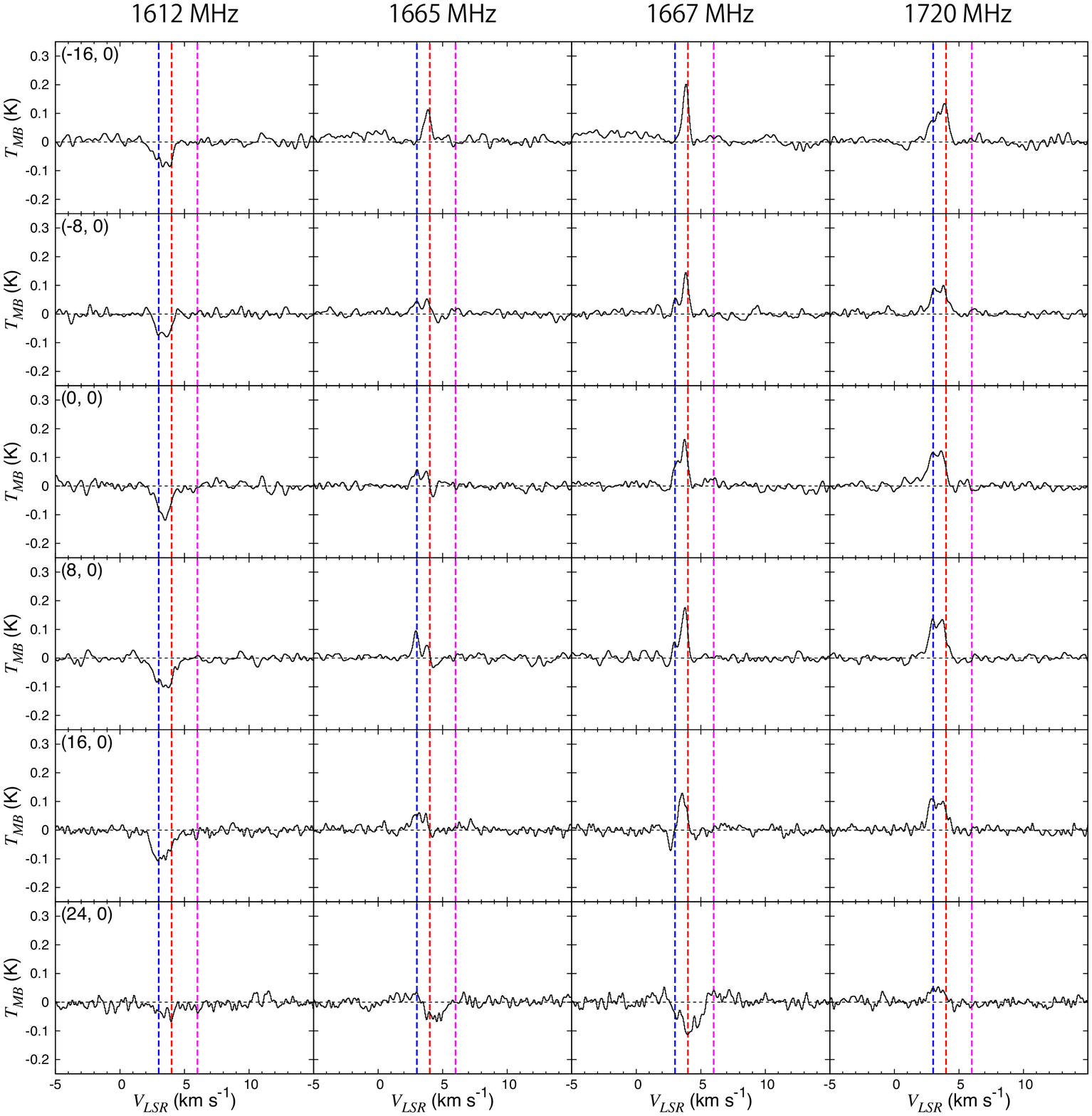}
    \caption{%
        Spectra of the four hfs components of the OH 18 cm transition
        observed toward the stem region of the Pipe nebula along strip-2 (Figure $\ref{fig:pipe_map_CO}$).
        The blue, red and magenta dotted lines represent the $V_{\rm LSR}$
        of 3.0, 4.0 and 6.0 km s$^{-1}$, respectively.
        \label{fig:pipe_spectra_strip2_stem}}
\end{figure}
\clearpage
\begin{figure}[!htbp]
    \centering
    \includegraphics[width=\hsize]{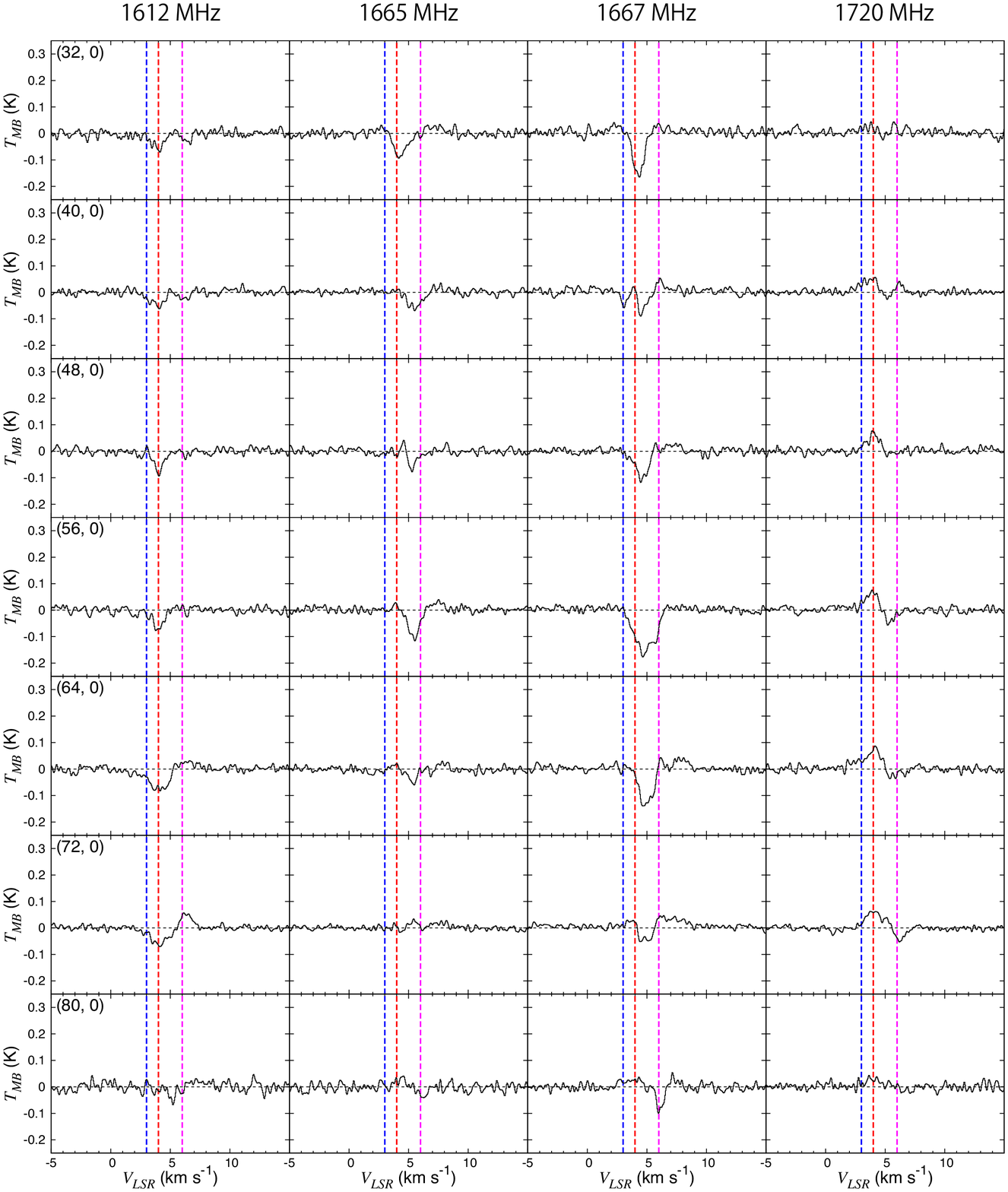}
    \caption{%
        Spectra of the four hfs components of the OH 18 cm transition
        observed toward the bowl region of the Pipe nebula along strip-2 (Figure $\ref{fig:pipe_map_CO}$).
        The blue, red and magenta dotted lines represent the $V_{\rm LSR}$
        of 3.0, 4.0 and 6.0 km s$^{-1}$, respectively.
        \label{fig:pipe_spectra_strip2_bowl}}
\end{figure}
\clearpage
\begin{figure}[!htbp]
    \centering
    \includegraphics[width=\hsize]{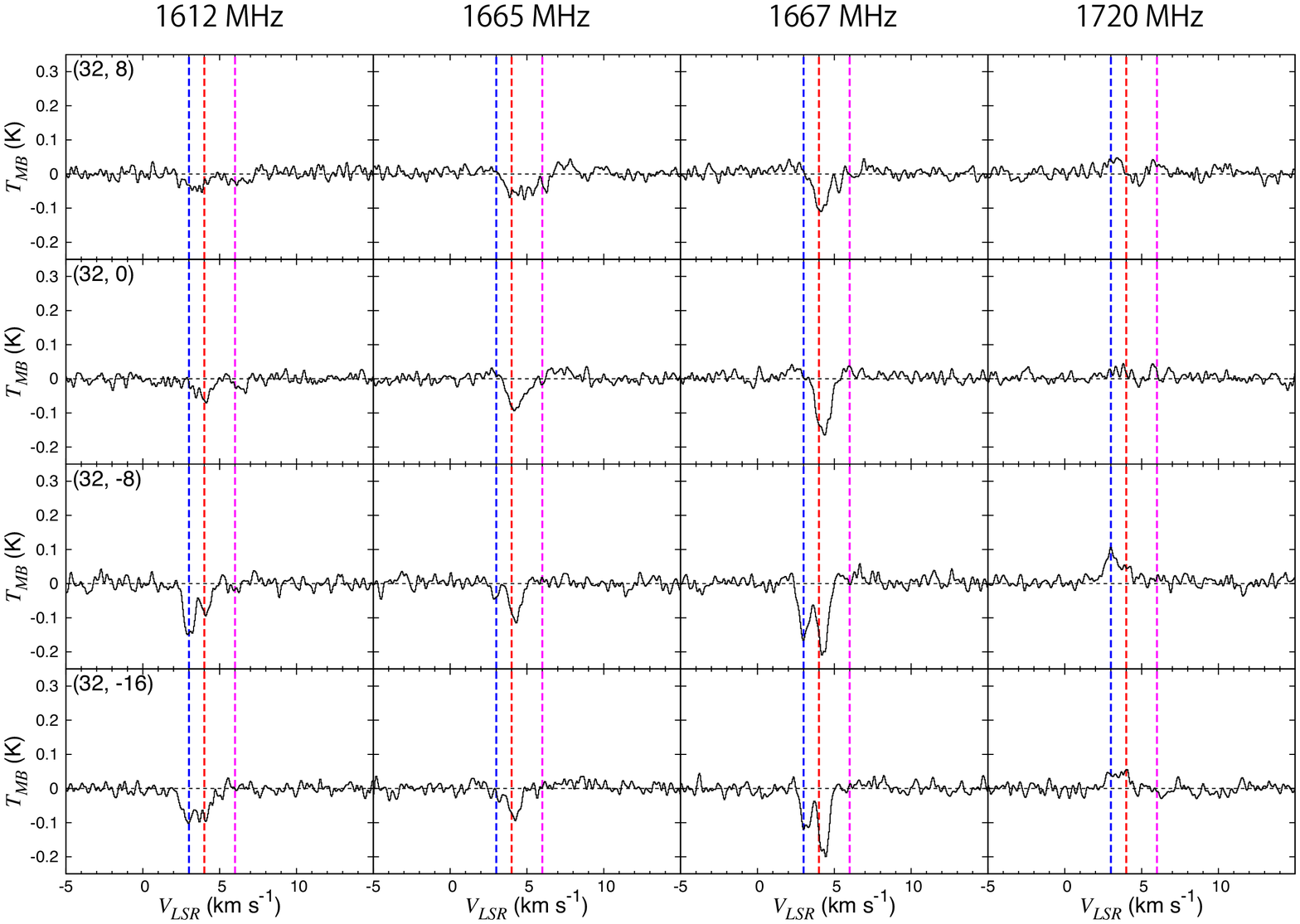}
    \caption{%
        Spectra of the four hfs components of the OH 18 cm transition
        observed toward the Pipe nebula along strip-3 (Figure $\ref{fig:pipe_map_CO}$).
        The blue, red and magenta dotted lines represent the $V_{\rm LSR}$
        of 3.0, 4.0 and 6.0 km s$^{-1}$, respectively.
        \label{fig:pipe_spectra_strip3}}
\end{figure}
\clearpage
\begin{figure}[!htbp]
    \centering
    \includegraphics[width=0.6\hsize]{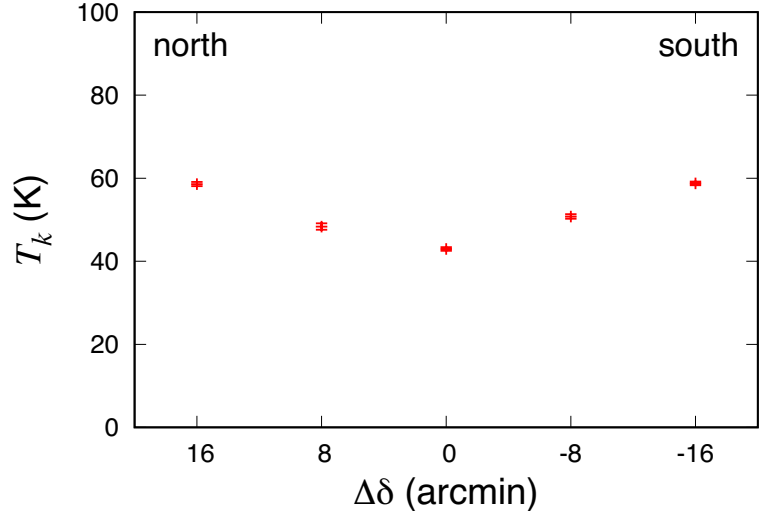}
    \caption{%
        The gas kinetic temperatures along strip-1 (Figure $\ref{fig:pipe_map_CO}$)
        determined by using our statistical equilibrium calculations.
        The abscissa is the angular offset from the center (Table $\ref{table:pipe_linepara_strip1}$).
        Error bars denote three times the standard deviation.
        \label{fig:pipe_Tk_strip1}}
\end{figure}
\clearpage
\begin{figure}[!htbp]
    \centering
    \includegraphics[width=0.7\hsize]{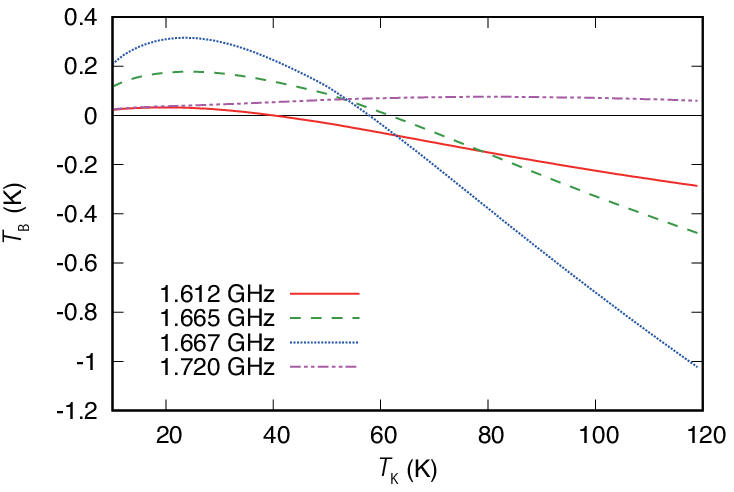}
    \caption{%
        The derived intensities of the four hfs lines of the OH 18 cm transition as a function
        of the gas kinetic temperature, where the H$_2$ density, the OH column density,
        and the H$_2$ ortho-to-para ratio are assumed to be
        10$^3$ cm$^{-3}$, 10$^{14}$ cm$^{-2}$, and 3, respectively.
        The intensity of the radio continuum background emision is assumed to be 5 K.
        \label{fig:LVG_mainabs_Tk}}
\end{figure}
\clearpage
\begin{figure}[!htbp]
    \centering
    \includegraphics[width=0.4\hsize]{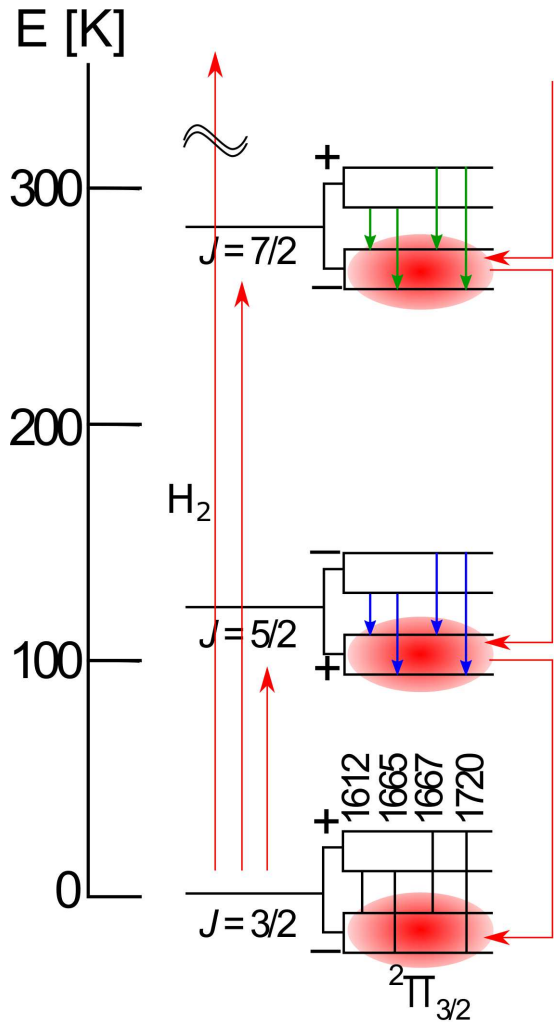}
    \caption{%
        The rotational energy diagram for the $^2\Pi_{3/2}$ states of the OH molecule.
        The overpopulation in the lower $\Lambda$-type doubling levels of the rotationally excited states
        is produced by the $\Lambda$-type doubling transitions
        (green and blue arrows).
        The subsequent rotational transitions (red downward arrows)
        produce the overpopulations in the $-$ levels of the $J$=3/2 states, respectively.
        \label{fig:OH_energy_mainabs}}
\end{figure}
\clearpage
\begin{figure}[!htbp]
    \centering
    \includegraphics[width=0.7\hsize]{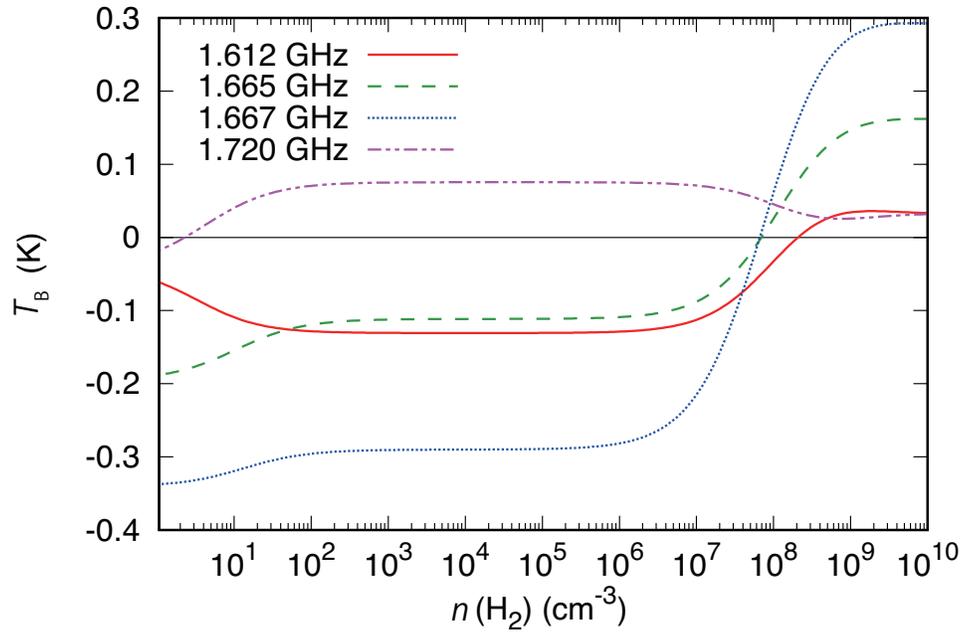}
    \caption{%
        The derived intensities of the four hfs lines of the OH 18 cm transition as a function
        of the H$_2$ density, where the OH column density, the gas kinetic temperature,
        and the H$_2$ ortho-to-para ratio are assumed to be
        10$^{14}$ cm$^{-2}$, 75 K and 3, respectively.
        The intensity of the radio continuum background emision is assumed to be 5 K.
        \label{fig:LVG_mainabs_nH2}}
\end{figure}
\clearpage
\begin{figure}[!htbp]
    \centering
    \includegraphics[width=\hsize]{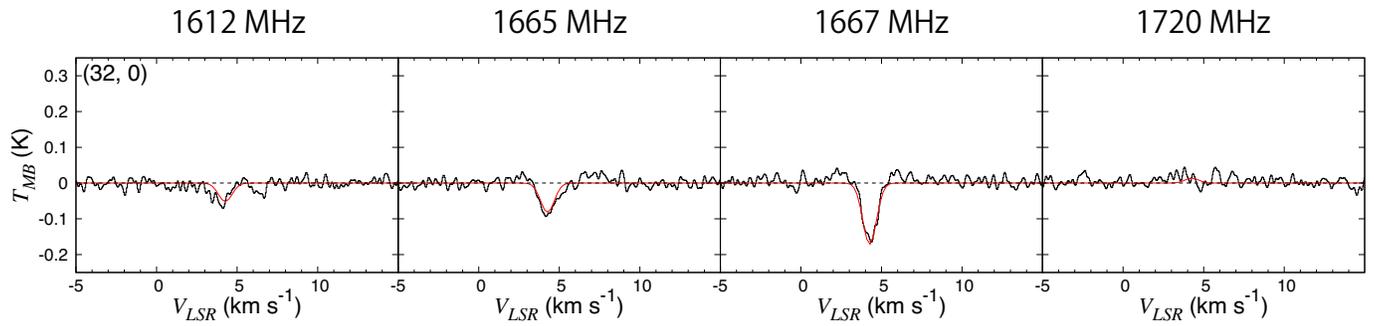}
    \caption{%
        Spectra for the four hfs components of the OH 18 cm transition
        observed on the (+32$'$, 0$'$) position on strip-2
        (Figure $\ref{fig:pipe_map_CO}$).
        Red lines show the Gaussian profiles with the best-fit parameters
        derived from our statistical equilibrium calculations,
        namely the gas kinetic temperature of 73.6 $\pm$ 0.8 K
        and the OH column density of (5.3 $\pm$ 0.2) $\times$ 10$^{13}$ cm$^{-2}$.
        \label{fig:pipe_spectra_fit}}
\end{figure}
\clearpage
\begin{figure}[!htbp]
    \centering
    \includegraphics[width=0.6\hsize]{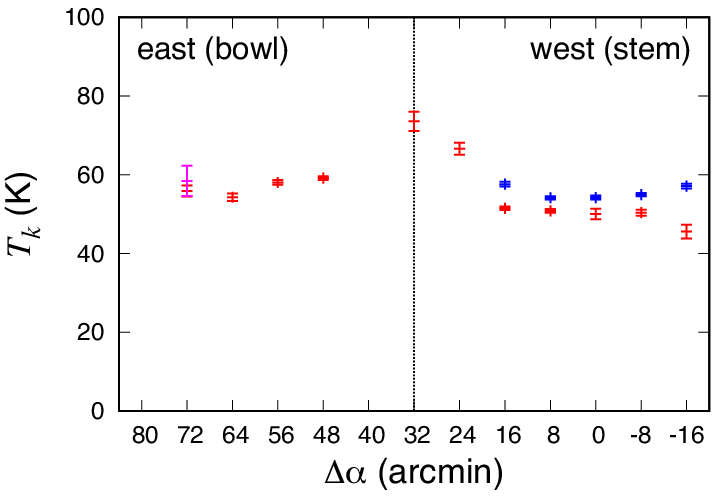}
    \caption{%
        The gas kinetic temperatures
        along strip-2 in Figure $\ref{fig:pipe_map_CO}$
        determined by using our statistical equilibrium calculations.
        The abscissa is the angular offset from the (0$'$, 0$'$) position
        (Table $\ref{table:pipe_linepara_strip2}$).
        Error bars denote three times the standard deviation.
        The blue, red and magenta color represent the $\sim$3, $\sim$4, $\sim$5--6 km s$^{-1}$
        velocity components, respectively.
        \label{fig:pipe_Tk_strip2}}
\end{figure}
\begin{figure}[!htbp]
    \centering
    \includegraphics[width=0.6\hsize]{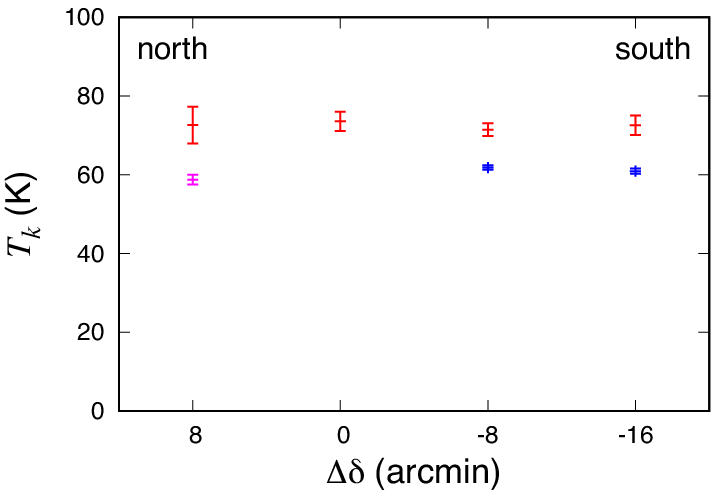}
    \caption{%
        The gas kinetic temperatures
        along strip-3 in Figure $\ref{fig:pipe_map_CO}$
        determined by using our statistical equilibrium calculations.
        The abscissa is the angular offset from the (32$'$, 0$'$) position
        (Table $\ref{table:pipe_linepara_strip3}$).
        Error bars denote three times the standard deviation.
        The blue, red and magenta color represent the $\sim$3, $\sim$4, $\sim$5--6 km s$^{-1}$
        velocity components, respectively.
        \label{fig:pipe_Tk_strip3}}
\end{figure}
\clearpage
\begin{table}[!htbp]
    \begin{center}
        \caption{%
            Observed line parameters toward the Pipe nebula along strip-1.
            \tablenotemark{a}
            \label{table:pipe_linepara_strip1}}
        \begin{tabular}{lllllll}
            \hline
            $\Delta\delta$ & $V_{\rm LSR}$     & FWHM          & $T_{\rm MB}$ (1612) & $T_{\rm MB}$ (1665) & $T_{\rm MB}$ (1667) & $T_{\rm MB}$ (1720) \\
            (arcmin)       & (km s$^{-1}$) & (km s$^{-1}$) & (K)             & (K)             & (k)             & (K)             \\ \hline
            -16            & 3.27 (2)      & 1.24 (4)      & -0.092 (3)      & -0.016 (3)      & 0.026 (3)       & 0.079 (3)       \\
            -8             & 3.216 (6)     & 0.82 (1)      & -0.073 (3)      & 0.055 (3)       & 0.192 (3)       & 0.105 (3)       \\
            0              & 3.298 (2)     & 0.568 (4)     & -0.074 (3)      & 0.226 (3)       & 0.450 (3)       & 0.188 (3)       \\
            8              & 3.336 (5)     & 0.82 (1)      & -0.063 (3)      & 0.088 (3)       & 0.229 (3)       & 0.080 (3)       \\
            16             & 3.33 (2)      & 1.30 (4)      & -0.061 (2)      & 0.005 (2)       & 0.051 (2)       & 0.053 (2)       \\
            \hline
        \end{tabular}
    \end{center}
    \tablenotetext{a}{
        $\Delta\delta$ in the first column denotes a Decl. offset from the
        ($\alpha_{2000}$, $\delta_{2000}$) = (17$^h$20$^m$49$^s$.0, -26$^\circ$53$'$8.0$''$) position.
        The main beam temperatures ($T_{\rm MB}$) of the four hfs lines of OH,
        the $V_{\rm LSR}$ value, and the linewidth (FWHM) are obtained by a Gaussian fit,
        assuming that the $V_{\rm LSR}$ values and the linewidths of the four hfs lines are
        identical for each observed position.
        The error in the parentheses represents one standard deviation,
        which applies to the last significant digits.
    }
\end{table}
\clearpage
\begin{table}[!htbp]
    \begin{center}
    \caption{%
        Observed line parameters toward the Pipe nebula along strip-2.
        \tablenotemark{a}
        \label{table:pipe_linepara_strip2}}
    \begin{tabular}{c|c|cccccc}
        \hline
                    & $\Delta\alpha$ & $V_{\rm LSR}$     & FWHM          & $T_{\rm MB}$ (1612) & $T_{\rm MB}$ (1665) & $T_{\rm MB}$ (1667) & $T_{\rm MB}$ (1720) \\
                    & (arcmin)       & (km s$^{-1}$) & (km s$^{-1}$) & (K)             & (K)             & (k)             & (K)             \\ \hline
        Stem region & -16            & 3.26 (4)      & 1.45 (6)      & -0.070 (4)      & 0.008 (3)       & 0.009 (4)       & 0.082 (4)       \\
                    &                & 3.84 (1)      & 0.54 (2)      & -0.036 (7)      & 0.111 (5)       & 0.190 (6)       & 0.086 (8)       \\
                    & -8             & 3.12 (2)      & 0.82 (4)      & -0.078 (3)      & 0.039 (3)       & 0.038 (3)       & 0.088 (3)       \\
                    &                & 3.85 (1)      & 0.50 (2)      & -0.063 (5)      & 0.041 (4)       & 0.134 (4)       & 0.089 (5)       \\
                    & 0              & 3.08 (3)      & 0.95 (6)      & -0.084 (4)      & 0.044 (3)       & 0.054 (4)       & 0.118 (4)       \\
                    &                & 3.74 (1)      & 0.57 (2)      & -0.083 (7)      & 0.022 (5)       & 0.138 (6)       & 0.082 (10)      \\
                    & 8              & 2.95 (1)      & 0.69 (2)      & -0.092 (4)      & 0.076 (3)       & 0.028 (4)       & 0.129 (4)       \\
                    &                & 3.73 (1)      & 0.63 (2)      & -0.114 (4)      & 0.023 (4)       & 0.160 (4)       & 0.132 (4)       \\
                    & 16             & 2.77 (1)      & 0.67 (3)      & -0.096 (4)      & 0.047 (4)       & -0.063 (5)      & 0.083 (4)       \\
                    &                & 3.55 (1)      & 0.89 (3)      & -0.096 (4)      & 0.038 (3)       & 0.114 (4)       & 0.102 (3)       \\
                    & 24             & 4.12 (2)      & 1.55 (5)      & -0.041 (3)      & -0.050 (3)      & -0.100 (3)      & 0.021 (3)       \\ \hline
        Bowl region & 32             & 4.26 (1)      & 1.01 (2)      & -0.056 (3)      & -0.095 (3)      & -0.162 (4)      & 0.015 (3)       \\
                    & 40             & -             & -             & -               & -               & -               & -               \\
                    & 48             & 4.27 (2)      & 1.33 (6)      & -0.066 (3)      & 0.011 (3)       & -0.084 (3)      & 0.063 (3)       \\
                    &                & 5.16 (1)      & 0.62 (3)      & 0.020 (5)       & -0.072 (4)      & -0.039 (6)      & -0.036 (5)      \\
                    & 56             & 4.14 (2)      & 1.33 (4)      & -0.069 (3)      & 0.032 (3)       & -0.095 (4)      & 0.078 (3)       \\
                    &                & 5.25 (2)      & 1.38 (4)      & 0.020 (3)       & -0.099 (3)      & -0.137 (3)      & -0.055 (3)      \\
                    & 64             & 4.22 (18)     & 1.69 (18)     & -0.10 (2)       & 0.027 (6)       & 0.007 (7)       & 0.10 (2)        \\
                    &                & 4.95 (2)      & 1.31 (5)      & 0.03 (3)        & -0.055 (9)      & -0.144 (5)      & -0.05 (3)       \\
                    & 72             & 3.79 (2)      & 1.23 (5)      & -0.06 (1)       & -0.003 (4)      & 0.03 (1)        & 0.058 (8)       \\
                    &                & 5.05 (4)      & 1.53 (17)     & -0.09 (6)       & -0.002 (13)     & -0.12 (7)       & 0.07 (5)        \\
                    &                & 5.70 (60)     & 2.27 (41)     & 0.07 (4)        & 0.015 (8)       & 0.08 (5)        & -0.06 (3)       \\
                    & 80             & 6.11 (2)      & 0.65 (4)      & -0.009 (5)      & -0.043 (5)      & -0.094 (5)      & -0.005 (4)      \\
         \hline 
    \end{tabular}
    \end{center}
    \tablenotetext{a}{
        $\Delta\alpha$ in the first column denotes a R.A. offset from the
        ($\alpha_{2000}$, $\delta_{2000}$) = (17$^h$27$^m$12$^s$.0, -26$^\circ$42$'$59.0$''$) position.
        The main beam temperatures ($T_{\rm MB}$) of the four hfs line of OH,
        the $V_{\rm LSR}$ value, and the linewidth (FWHM) are obtained by a Gaussian fit,
        assuming that the $V_{\rm LSR}$ values and the linewidths of the four hfs lines are
        identical for each observed position.
        The error in the parentheses represents one standard deviation,
        which applies to the last significant digits.
    }
\end{table}
\clearpage
\begin{table}[!htbp]
    \begin{center}
    \caption{%
        Observed line parameters toward the Pipe nebula along strip-3.
        \tablenotemark{a}
        \label{table:pipe_linepara_strip3}}
    \begin{tabular}{c|cccccc}
        \hline
        $\Delta\delta$ & $V_{\rm LSR}$     & FWHM          & $T_{\rm MB}$ (1612) & $T_{\rm MB}$ (1665) & $T_{\rm MB}$ (1667) & $T_{\rm MB}$ (1720) \\
        (arcmin)       & (km s$^{-1}$) & (km s$^{-1}$) & (K)             & (K)             & (k)             & (K)             \\ \hline
        8              & 4.16 (2)      & 1.20 (5)      & -0.033 (3)      & -0.065 (3)      & -0.097 (4)      & 0.002 (3)       \\
                       & 5.84 (4)      & 1.2 (1)       & -0.025 (3)      & -0.042 (4)      & -0.004 (3)      & 0.026 (3)       \\
        0              & 4.26 (1)      & 1.01 (2)      & -0.056 (3)      & -0.095 (3)      & -0.162 (4)      & 0.015 (3)       \\
        -8             & 3.05 (1)      & 0.79 (2)      & -0.160 (4)      & -0.031 (4)      & -0.155 (4)      & 0.103 (4)       \\
                       & 4.24 (1)      & 0.75 (2)      & -0.080 (4)      & -0.112 (4)      & -0.210 (5)      & 0.047 (4)       \\
        -16            & 3.21 (2)      & 1.11 (5)      & -0.102 (4)      & -0.028 (3)      & -0.094 (4)      & 0.049 (3)       \\
                       & 4.32 (1)      & 0.68 (2)      & -0.065 (5)      & -0.090 (4)      & -0.198 (5)      & 0.021 (4)       \\
         \hline 
    \end{tabular}
    \end{center}
    \tablenotetext{a}{
        $\Delta\delta$ in the first column denotes a Decl. offset from the
        the (32$'$, 0$'$) position on strip-2 (Table $\ref{table:pipe_linepara_strip2}$).
        The main beam temperatures ($T_{\rm MB}$) of the four hfs line of OH,
        the $V_{\rm LSR}$ value, and the linewidth (FWHM) are obtained by a Gaussian fit,
        assuming that the $V_{\rm LSR}$ values and the linewidths of the four hfs lines are
        identical for each observed position.
        The error in the parentheses represents one standard deviation,
        which applies to the last significant digits.
    }
\end{table}
\clearpage
\begin{table}[!htbp]
    \begin{center}
        \caption{%
            The derived parameters along strip-1.\tablenotemark{a}
            \label{table:pipe_Tk_strip1}}
        \begin{tabular}{c|cc|c}
            \hline
            $\Delta\delta$ (arcmin) & $T_k$ (K) & $N$(OH) (10$^{14}$ cm$^{-2}$) & $T_{\rm cont}$(1667) (K) \\ \hline
            16                      & 58.6 (2)  & 1.13 (2)                      & 4.5                  \\
            8                       & 48.3 (3)  & 1.00 (1)                      & 4.6                  \\
            0                       & 42.9 (1)  & 1.178 (8)                     & 4.7                  \\
            -8                      & 50.8 (2)  & 1.02 (1)                      & 4.8                  \\
            -16                     & 58.8 (1)  & 1.27 (2)                      & 4.9                  \\
            \hline 
        \end{tabular}
    \end{center}
    \tablenotetext{a}{%
        $\Delta\delta$ in the first column denotes a Decl. offset from the
        ($\alpha_{2000}$, $\delta_{2000}$) = (17$^h$20$^m$49$^s$.0, -26$^\circ$53$'$8.0$''$) position.
        The gas kinetic temperature ($T_k$) and the OH column density ($N$(OH))
        are determined from our statistical equilibrium calculation
        assuming the density and the ortho-to-para ratio of H$_2$ of 10$^3$ cm$^{-3}$ and 3, respectively.
        The numbers in the parentheses represent one standard deviation of the fit
        in units of the last significant digits.
        $T_{\rm cont}$ in the last column is the intensity of the radio continuum background emission
        at the frequency of 1667 MHz evaluated from the HIPASS 1.4 GHz data \citep{Calabretta2014}.
    }
\end{table}
\clearpage
\begin{table}[!htbp]
    \begin{center}
        \caption{%
            The derived parameters along strip-2.\tablenotemark{a}
            \label{table:pipe_Tk_strip2}}
        \begin{tabular}{c|cc|cc|c}
            \hline
                        & $\Delta\alpha$ (arcmin) & $V_{\rm LSR}$ (km s$^{-1}$) & $T_k$ (K) & $N$(OH) (10$^{14}$ cm$^{-2}$) & $T_{\rm cont}$(1667) (K) \\ \hline
            Stem region & -16                     & 3.26                    & 57.1 (2)  & 1.47 (3)                      & 5.2                  \\
                        &                         & 3.84                    & 45.6 (6)  & 0.63 (2)                      & 5.2                  \\
                        & -8                      & 3.13                    & 55.0 (1)  & 0.93 (1)                      & 5.2                  \\
                        &                         & 3.85                    & 50.3 (3)  & 0.54 (1)                      & 5.2                  \\
                        & 0                       & 3.08                    & 54.2 (2)  & 1.19 (2)                      & 5.3                  \\
                        &                         & 3.74                    & 49.8 (5)  & 0.56 (3)                      & 5.3                  \\
                        & 8                       & 2.95                    & 54.2 (1)  & 0.95 (1)                      & 5.3                  \\
                        &                         & 3.73                    & 50.8 (2)  & 0.88 (1)                      & 5.3                  \\
                        & 16                      & 2.77                    & 57.6 (2)  & 0.79 (1)                      & 5.4                  \\
                        &                         & 3.55                    & 51.5 (2)  & 1.11 (2)                      & 5.4                  \\
                        & 24                      & 4.12                    & 66.6 (5)  & 0.85 (3)                      & 5.5                  \\ \hline
            Bowl region & 32                      & 4.26                    & 73.6 (8)  & 0.53 (2)                      & 5.5                  \\
                        & 40                      & -                       & -         & -                             & 5.6                  \\
                        & 48                      & 4.27                    & 59.2 (2)  & 1.25 (2)                      & 5.6                  \\
                        &                         & 5.16                    & -         & -                             & 5.6                  \\
                        & 56                      & 4.14                    & 58.1 (2)  & 1.35 (2)                      & 5.7                  \\
                        &                         & 5.25                    & -         & -                             & 5.7                  \\
                        & 64                      & 4.22                    & 54.3 (3)  & 1.9 (1)                       & 5.7                  \\
                        &                         & 4.95                    & -         & -                             & 5.7                  \\
                        & 72                      & 3.79                    & 55.9 (5)  & 0.93 (7)                      & 5.7                  \\
                        &                         & 5.05                    & 58 (1)    & 1.6 (3)                       & 5.7                  \\
                        &                         & 5.70                    & -         & -                             & 5.7                  \\
                        & 80                      & 6.11                    & -         & -                             & 5.8                  \\
             \hline 
        \end{tabular}
    \end{center}
    \tablenotetext{a}{%
        $\Delta\alpha$ in the first column denotes a R.A. offset from the
        ($\alpha_{2000}$, $\delta_{2000}$) = (17$^h$27$^m$12$^s$.0, -26$^\circ$42$'$59.0$''$) position.
        The gas kinetic temperature ($T_k$) and the OH column density ($N$(OH))
        are determined from our statistical equilibrium calculation
        assuming the density and the ortho-to-para ratio of H$_2$ of 10$^3$ cm$^{-3}$ and 3, respectively.
        The numbers in the parentheses represent one standard deviation of the fit
        in the units of the last significant digits.
        $T_{\rm cont}$ in the last column is the intensity of the radio continuum background emission
        at the frequency of 1667 MHz evaluated from the HIPASS 1.4 GHz data \citep{Calabretta2014}.
    }
\end{table}
\clearpage
\begin{table}[!htbp]
    \begin{center}
        \caption{%
            The derived parameters along strip-3.\tablenotemark{a}
            \label{table:pipe_Tk_strip3}}
        \begin{tabular}{cc|cc|c}
            \hline
            $\Delta\delta$ & $V_{\rm LSR}$ (km s$^{-1}$) & $T_k$ (K) & $N$(OH) (10$^{14}$ cm$^{-2}$) & $T_{\rm cont}$(1667) (K) \\ \hline
            8              & 4.16                    & 73 (5)    & 0.43 (4)                      & 5.5                  \\
                           & 5.84                    & 59 (1)    & 0.55 (4)                      & 5.5                  \\
            0              & 4.26                    & 74 (2)    & 0.53 (2)                      & 5.5                  \\
            -8             & 3.05                    & 62 (1)    & 1.00 (1)                      & 5.6                  \\
                           & 4.24                    & 71 (2)    & 0.55 (2)                      & 5.6                  \\
            -16            & 3.21                    & 61 (1)    & 1.03 (2)                      & 5.6                  \\
                           & 4.32                    & 73 (2)    & 0.41 (2)                      & 5.6                  \\
            \hline 
        \end{tabular}
    \end{center}
    \tablenotetext{a}{%
        $\Delta\delta$ in the first column denotes a Decl. offset from the
        the (32$'$, 0$'$) position on strip-2 (Table $\ref{table:pipe_linepara_strip2}$).
        The gas kinetic temperature ($T_k$) and the OH column density ($N$(OH))
        are determined from our statistical equilibrium calculation
        assuming the density and the ortho-to-para ratio of H$_2$ of 10$^3$ cm$^{-3}$ and 3, respectively.
        The numbers in the parentheses represent one standard deviation of the fit
        in the units of the last significant digits.
        $T_{\rm cont}$ in the last column is the intensity of the radio continuum background emission
        at the frequency of 1667 MHz evaluated from the HIPASS 1.4 GHz data \citep{Calabretta2014}.
    }
\end{table}
\end{document}